\documentclass{gGAF2e}

\newcommand{\down}[1]{\mathrm{#1}}
\newcommand{\myvec}[1]{\mathbf{#1}}

\usepackage{natbib}

\begin{document}

\doi{10.1080/03091920xxxxxxxxx}
\issn{1029-0419} \issnp{0309-1929} \jvol{00} \jnum{00} \jyear{2009}
\jmonth{Month}

\markboth{Heyner et al.}{The initial temporal evolution of a feedback dynamo for
Mercury}

\title{The initial temporal evolution of a feedback dynamo for Mercury}

\author{D. HEYNER$\dagger$$^{\ast}$\thanks{$^\ast$Corresponding author. Email:
d.heyner@tu-bs.de\vspace{6pt}},
D. SCHMITT$\ddagger$,
J. WICHT$\ddagger$,
K.-H. GLASSMEIER$\dagger\ddagger$,
H. KORTH$\S$,
and U. MOTSCHMANN$\P\|$
\\\vspace{6pt}
$\dagger$TU Braunschweig, Institut f\"ur Geophysik und extraterrestrische
Physik, Mendelssohnstr. 3, 38106 Braunschweig, Germany\\
$\ddagger$Max-Planck-Institut f\"ur Sonnensystemforschung, Max-Planck-Str. 2,
37191 Katlenburg-Lindau, Germany\\
$\S$The Johns Hopkins University, Applied Physics Laboratory, 11100 Johns
Hopkins Rd, Laurel MD 20723-6099, USA\\
$\P$TU Braunschweig, Institut f\"ur Theoretische Physik, Mendelssohnstr. 3,
38106 Braunschweig, Germany\\
$\|$DLR Institut f\"ur Planetenforschung, 12489 Berlin, Germany
}

\maketitle

\begin{abstract}
Various possibilities are currently under discussion to explain the
observed weakness of the intrinsic magnetic field of planet Mercury. One of the
possible dynamo scenarios is a dynamo with feedback from the magnetosphere. Due
to its weak magnetic field Mercury exhibits a small magnetosphere whose
subsolar magnetopause distance is only about 1.7 Hermean radii. We
consider the magnetic field due to magnetopause currents in the dynamo region.
Since the external field of magnetospheric origin is antiparallel to the dipole
component of the dynamo field, a negative feedback results. For an
$\alpha\Omega$-dynamo two stationary solutions of such a feedback dynamo
emerge, one with a weak and the other with a strong magnetic field. The
question, however, is how these solutions can be realized. To address
this problem, we discuss various scenarios for a simple dynamo model and the
conditions under which a steady weak magnetic field can be reached. We
find that the feedback mechanism quenches the overall field to a low value of
about 100 to 150 nT if the dynamo is not driven too strongly.\bigskip
\bigskip
\end{abstract}

\section{Introduction}

The recent flybys of the MESSENGER spacecraft at planet Mercury confirm the
existence of a large scale magnetic field \citep{messenger_2009}. The
dipole surface field, however, is roughly one to two orders of magnitude too
weak to be commensurable with classical dynamo theory \citep{wicht_2007,
olson_christensen_2006}.  There are several approaches to explain this
disagreement \citep{Heimpel_2005, Stanley_2005, Christensen_2006,
Matsushima_2006, glassmeier_2007} with different dynamo configurations. Here,
we further study the feedback dynamo scenario suggested by
\citet{glassmeier_2007} who investigated the interaction of the dynamo and the
magnetospheric field. They derived two stationary solutions and ascribed
the weaker solution to Mercury's magnetic field. They however do not address
the question how the dynamo reaches either of these solutions. Allowing a
variable magnetopause which depends on the internal field and solar wind
conditions, it is so far not conceivable how a dynamo can develope into a state
where it can be quenched by the external feedback field. Therefore, the present
study aims at discussing conditions under which a steady and weak magnetic
field can evolve when the dynamo is exposed to a magnetospheric magnetic field.

\section{A Hermean feedback dynamo}

The magnetopause currents caused by the interaction of Mercury's magnetic field
with the solar wind generate an external field which reaches into the planet's
interior.  Since the internal magnetic field is weak, the magnetopause is
located close to the planet. We thus expect a stronger external field
contribution in the dynamo region than, for example, for Earth. In the
terrestrial case, the subsolar magnetopause is located at about 10 planetary
radii and its influence on the internal dynamo process is negligible.  In
contrast, at Mercury the magnetopause is close to the planet and the
external field has to be taken into account in the solution of the dynamo
problem.\\

\begin{figure}
 \centering\includegraphics[width=8cm]{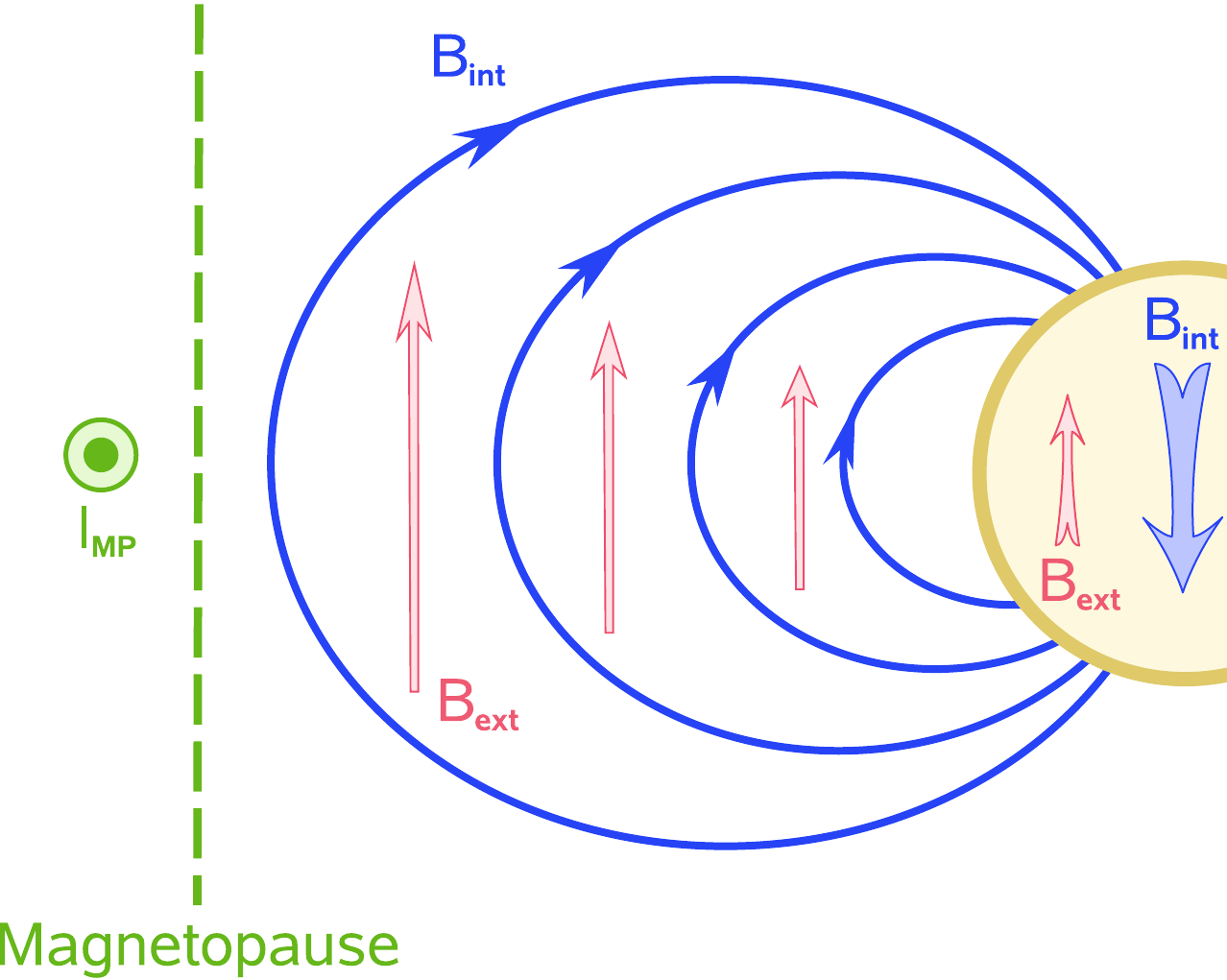}
    \caption{Schematic illustration of the feedback mechanism. The planet's
    dynamo generates an internal field $B_\down{int}$. The interaction with the
    solar wind causes a magnetopause current $I_\down{MP}$ which itself induces
    an external field $B_\down{ext}$ which is of opposite orientation to the
    internal field in the Hermean core \citep{glassmeier_2007}.}
    \label{fig:external_field_topology}
\end{figure}

The relative orientation of the internal and external magnetic fields is of
significance.  As seen in figure \ref{fig:external_field_topology}, the
magnetopause currents generate a field canceling the field outside the
magnetosphere.  Inside the magnetopause internal and external fields are
parallel.  At the core-mantle boundary the situation is different.  The
internal dipole field possesses a vector component along the rotational axis of
the planet that is anti-parallel to the externally generated magnetic field.
Thus, a negative feedback situation results.\\

Since the feedback field is stronger for a close magnetopause, we concentrate
on a weak initial dynamo field.  This situation corresponds to the onset of
dynamo action or to the time period after a polarity reversal when the dipole
field is weak compared to higher multipoles.  In order to gain first insights
into the system's possible temporal evolution, we reduce its complexity by
coupling a simple kinematic internal dynamo to an idealized external
magnetospheric field.

\section{Response function}
The external field arising from magnetopause currents depends on the distance
of the magnetopause to the planet and the spatial current distribution. The
dynamical magnetopause position, parameterized by the stand-off distance $R_s$
at the subsolar point, is mainly determined by the pressure equilibrium between
the planetary magnetic pressure, with the dipolar part as the main
contribution and the solar wind dynamic pressure
\citep[e.g.][]{basic_space_plasma_physics_1996}:
\begin{equation}
   \label{eq:pressure_balance}
   R_s (g_{1,\down{int}}^0)= R_M \left(\frac{2 \; (g_{1,\down{int}}^0)^2}{\mu_0
   p_\down{sw}}\right)^{1/6} \quad .
\end{equation}
Here $R_M$, $\mu_0$, $g_{1,\down{int}}^0$ and $p_\down{sw}$ denote the Hermean
planetary radius, the permeability of free space, the internal axial dipole
Gauss coefficient and the solar wind ram pressure, respectively. Equation
(\ref{eq:pressure_balance}) demonstrates that the stand-off distance depends on
the internal field strength like $(g_{1,\down{int}}^0)^{1/3}$. The magnetopause
is thus located close to the planet for weak magnetic fields like the one found
at Mercury. In contrast to that, Earth exhibits a relatively strong magnetic
field with a distant magnetopause and negligible influence on the
internal dynamics in the planet's core.\\

In general, when the shape of the magnetopause and the planetary dipole
field strength are given and the solar-wind is assumed to be field-free, the
external field from magnetopause currents can be calculated without explicitly
determining the currents.  This is achieved by shielding the internal field by
an external potential field at the magnetopause, such that the magnetic flux
through the magnetopause vanishes. The field-free approximation is applied
since incorporating the ever-changing interplanetary magnetic field (IMF)
characteristics would require detailed hybrid modeling of the solar wind
interacting with the planetary magnetic field or long-term in-situ magnetic
field observations which are not available at this time. Altogether, in order
to determine the external field the stand-off distance must be known and then
the internal dipole field strength sets the shielding current strength. It is
therefore possible to express the external field strength $B_\down{ext}$ as a
response function $f$ of the internal one maintained by the dynamo process:
\begin{equation}
   \label{eq:allg_feedback_function}
   B_\down{ext} = f \left(g_{1,\down{int}}^0\right) \quad .
\end{equation}
There exist several models for the external field for various solar wind and
planetary magnetic field conditions.  For the Hermean case,
\citet{glassmeier_2007} constructed a simple model treating the
magnetopause current as a circular line current in the equatorial plane. The
well-studied terrestrial situation can be described with a semi-empirical model
by \citet{Tsyganenko_2005}. In that study, the stand-off distance and the
current strength depend on solar wind conditions. Making use of the
aforementioned field-free approximation, the model prescribes the
magnetopause shape as an ellipsoid with a cylindrical continuation as a
magnetotail. The spatial parameters of this magnetopause are fitted to
satellite observations. At this boundary the magnetic field of the planet
represented by its dipolar part and contributions arising from several
magnetospheric current systems are partially shielded depending on IMF
conditions.\\

For a more realistic representation than \citet{glassmeier_2007} we adapt the
terrestrial model of Tsyganenko to Hermean conditions following the approach of
\citet{korth_2004}. First, we assume a centered axial dipole as the planet's
intrinsic magnetic field.  Furthermore, since there are no permanently trapped
particles expected because of the low internal field strength of Mercury, we
neglect the magnetospheric ring current.  In order to scale this Tsyganenko
model to Hermean conditions we make use of the scaling law
\begin{equation}
   \myvec{B}_\down{M} (\myvec{r}_\down{M}) = \myvec{B}_\down{E}
(\myvec{r}_\down{E})
   = \myvec{B}_\down{E} (\kappa \, \myvec{r}_\down{M}) \quad \mbox{with} \quad
\kappa
   = \kappa_p \; \kappa_B =
   \left(\frac{p_\down{sw,M}}{2\,\mbox{nPa}}\right)^{0.14}\left( \frac{30,574\,
   \mbox{nT}}{g^0_{1, \down{int,M}}}\right)^{1/3} \quad .
\end{equation}
Here $\myvec{B}_\down{M}(\myvec{r}_\down{M})$,
$\myvec{B}_\down{E}(\myvec{r}_\down{E})$ and $g^0_{1, \down{int,M}}$ denote a
magnetic field vector in the Hermean system at the position
$\myvec{r}_\down{M}$, a magnetic field vector in the terrestrial system at the
position $\myvec{r}_\down{E}$ and the internal, axisymmetric dipole
Gauss-coefficient of the magnetic potential expansion.   The scaling factor
$\kappa_p$ due to
different solar wind ram pressures at Earth and Mercury has been
extrapolated from a
model by
\citet{Tsy_1996} fitting satellite observations at different solar wind
conditions
to their model magnetopause. At
Mercury the solar wind ram pressure is taken to be $13.4$ nPa assuming an
average solar wind speed of $400$ km/s and an average proton number density of
$5 \times 10^7 \mbox{ m}^{-3}$ \citep[e.g.][]{glassmeier_1997}.  The value of
$2$ nPa is
the average solar wind ram pressure at Earth. The second
factor $\kappa_B$, accounting for the different magnetic moments of the
planets, has already been used by \citet{korth_2004}.  The factor $30,574$ nT is
the
terrestrial magnetic dipole field strength at the equator around 1980 as it has
been used in the \citet{Tsy_1996} model. The linear factor $\kappa_B$ thus
scales the magnetospheres in such a way that the equatorial field strengths in
both planetary systems are equal.\\

The described magnetospheric model provides the full spectral multipole
information of the external field in response to any internal dipole field
strength. The dynamo, however, is only affected by a long-term average
magnetopause field. The time span needed for an external field to diffuse
through the entire core region is of the order of
\begin{equation}
   \label{eq:diffusion_time}
   \tau = \frac{L^2}{\eta} = \mu_0 \sigma L^2 \approx 35 \times 10^3 \mbox{a}
   \; ,
\end{equation}
where $\eta$ is the magnetic diffusivity, $\sigma = 6 \times 10^5$ S/m is the
assumed electrical conductivity \citep{suess_1979} and $L = R_\down{cmb} -
R_\down{icb}$ is the radius of the outer core shell with $R_\down{cmb}=1860$ km
as the radius of the core-mantle boundary \citep{spohn_2001}. The inner core
radius $R_\down{icb}$ is not well constrained but an Earth-like value of $0.35
R_\down{cmb}$ is chosen here. Therefore, any external non-axisymmetric magnetic
multipole contribution to the overall field in the planet's interior would
cancel over the Hermean orbital rotation period (88 Earth-days) and planetary
rotation period (59 Earth-days).\\

\begin{figure}
 \centering\includegraphics[width=8cm]{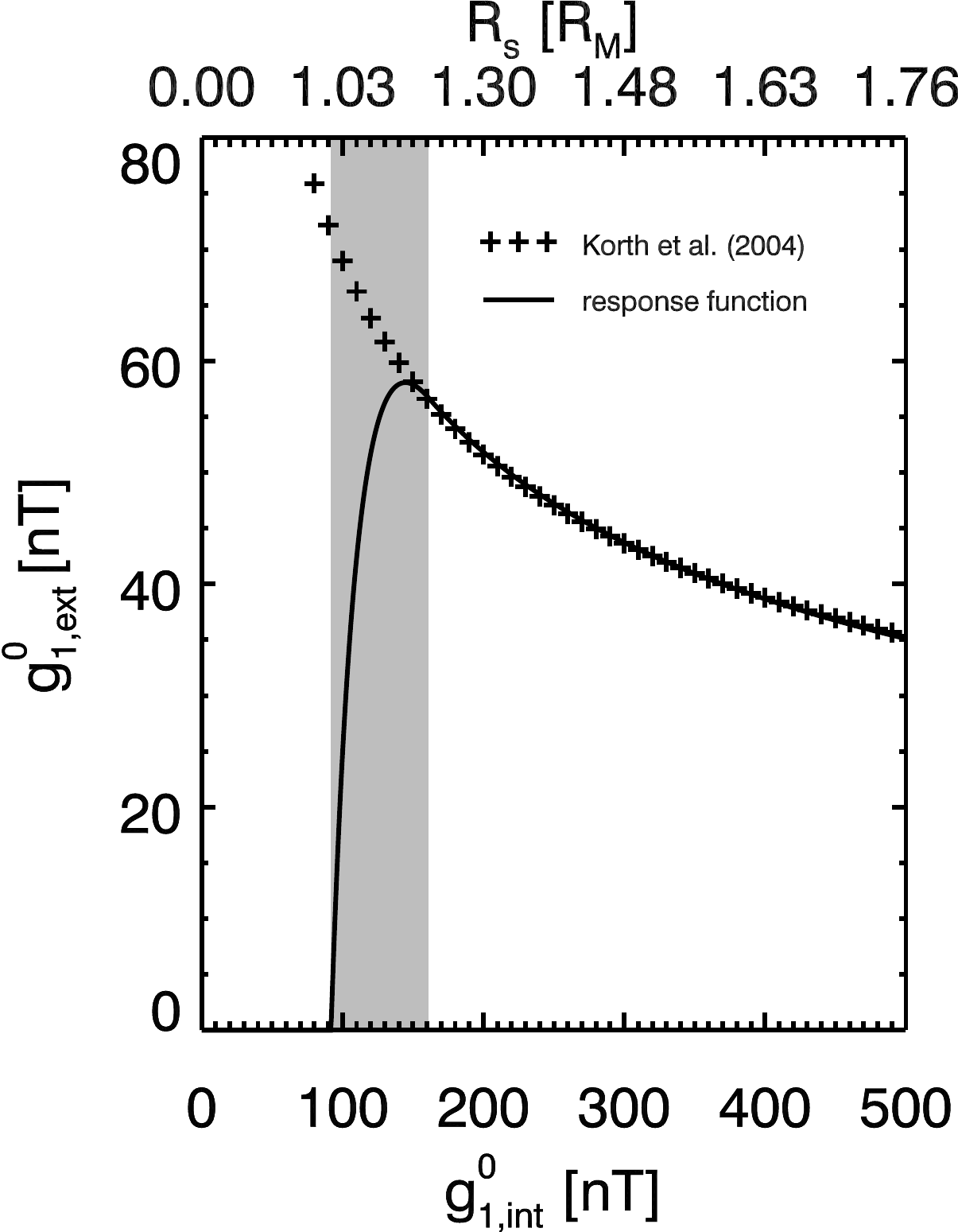}
    \caption{External dipole field strength as a function of the internal
    dipole field strength. The crosses indicate the values computed with
    the scaled Tsyganenko model and the solid line marks the model function with
    incorporated transition. The greyed area visualizes this transition region
    where the mode of interaction between the solar wind and planet is changed
    due to the planetary surface.}
    \label{fig:feedback_function}
\end{figure}

As external multipoles of degree $l > 1$ decrease towards the planet, we
furthermore restrict ourselves for simplicity reasons to the strongest
multipole $l=1$. This provides a uniform magnetic field in the interior
which is aligned with the planetary rotation axis. As an estimate of this
external field we take the magnetic field value obtained from the model at the
subsolar point on the Hermean equator. The resulting hyperbolic response
function is shown by crosses in figure \ref{fig:feedback_function}, where
the external and internal fields are represented by their multipole
coefficients $g^0_\down{1,ext}$ and $g^0_\down{1,int}$, respectively. For a
strong internal field, the magnetopause is pushed further away from the planet,
thus resulting in a weak external feedback.  In contrast, the external feedback
is strong for weak internal fields since the magnetopause is closer to the
planetary surface.

We modify the response function to exclude the unrealistic case of a stand-off
distance located within the planet, $R_s\le R_M$, which is equivalent to
$g^0_\down{1,int} \leq 91.8 $ nT according to (\ref{eq:pressure_balance}).
Furthermore, we need to take into account the finite extent of the magnetopause
\citep{russell_1982}. This implies that the magnetopause currents are
distributed over a finite radial extent. Some of the consequences for magnetic
field measurements in the Hermean system are discussed by
\citet{bepi_pss_2009}. While detailed modeling of the complex magnetopause
current structure is beyond the scope of the present study, we take a first
step to respect the finite thickness. The full response function is assumed to
apply only at distances greater than 500 km away from the planetary surface.
This corresponds to a typical magnetopause thickness \citep{russell_1982}.  In
consequence, the planetary dipole coefficient must exceed about $161$ nT to
yield an entirely undisturbed magnetopause. For weaker fields, i.e., for
smaller $R_\down{s}$, we modify the response function such that a smooth
transition towards the $R_\down{s} \leq R_\down{M}$ situation emerges. The
solid line in figure \ref{fig:feedback_function} shows this modified feedback
function, whose functional form is described in the following.  Throughout the
transition interval $91.8 \mbox{ nT } \leq g_{1,\down{int}}^0 \leq 161$ nT we
model the external field with a response function
\begin{equation}
g_{1, \down{ext}}^0 = 9.00\times 10^6 \; (g_{1, \down{int}}^0 - 91.8 \mbox{ nT})
\;
   (g_{1, \down{int}}^0 /\mbox{nT})^{-2.73} ,
\end{equation}
while for the remaining interval $g_{1,\down{int}}^0 > 161$ nT the feedback
function is parameterized with an exponential function
\begin{equation}
   g_{1, \down{ext}}^0 =  1.37 \times 10^5 \mbox{ nT }\; \exp\left(-5.96
\; (g_{1, \down{int}}^0 / \mbox{nT})^{5.27\times 10^{-2}} \right)
\end{equation}
fitted to the findings from the Tsyganenko model.
This parameterized response function allows us to calculate the influence of
the magnetospheric magnetic field on the dynamo without explicitly evaluating
the stand-off distance and the modified Tsyganenko model.

\section{An $\alpha\Omega$-dynamo embedded in an external magnetic field}

In order to describe the influence of an imposed external magnetic field
on the dynamo process, an additional induction term in the dynamo equation is
introduced \citep{levy_1979,glassmeier_2007}:
\begin{equation}
   \label{eq:normal_induction_equation}
   \frac{\partial \myvec{B}}{\partial t} = \myvec{\nabla} \times \left[
\myvec{v}
   \times \left( \myvec{B} + \myvec{B}_\down{ext} \right) \right] + \eta
   \Delta \myvec{B}
\end{equation}
where $\myvec{v}$ denotes the velocity, $\eta$ the magnetic diffusivity,
$\myvec{B}$ the dynamo field and $\myvec{B}_\down{ext}$ the external
magnetic field. To study the temporal evolution of the feedback dynamo we
adapt a version of a 1D kinematic mean-field $\alpha\Omega$-model presented by
\citet{schmitt_1989}, who studied different non-linear quenching mechanisms with
application to the Sun. With the magnetospheric feedback we introduce
another non-linear quenching method but within the context of a planet with a
magnetosphere. The main scope of this paper is to address the question how the
coupled system can dynamically evolve into a weak field solution. The simple
 kinematic dynamo serves to get a first picture of the various
scenarios that may arise.\\

The model considers dynamo action in a differentially rotating
spherical shell with an outer core radius of $R_\down{cmb}$. The radial
variation of the magnetic field and of the induction effects are neglected.
About the latter little is known in the case of Mercury and any specification
seems arbitrary. The neglect of a radial dependence of the magnetic field is
only permitted for a thick shell, as it is probably the case for Mercury's
fluid core. Furthermore, we assume rotational symmetry, so that all quantities
are independent of the azimuth ($\partial_\varphi = 0$) and thus depend solely
on the colatitude $\theta$. The magnetic field is decomposed into a
poloidal component, described by a vector potential $\myvec{A} = (0,0,A)$ and a
toroidal magnetic field component $(0,0,B)$ by
\begin{equation}
   \myvec{B}= (0,0,B) + \myvec{\nabla} \times (0,0,A) \quad .
\end{equation}
The toroidal field is produced by a constant radial shear $\partial_r \Omega =
\Omega'= \mbox{const.}$ through the so-called $\Omega$-effect. The poloidal
field is maintained by the $\alpha$-effect, a parameterized interaction of
small-scale field and small-scale flow. We assume a simple harmonic dependence
$\alpha(\theta) = \alpha_0 \cos\theta$.\\

In order to compute the magnetic field in the magnetosphere, the field must be
continued outside the dynamo shell.  Since the radial component of the magnetic
field must be continuous at the core-mantle boundary we can analyze
$B_r = (\myvec{\nabla} \times \myvec{A})_r$
to find the internal dipole Gauss coefficient $g_{1,\down{int}}^0$.  Any
possible influence of the embedding electrically conducting mantle is neglected
here. With the internal dipole coefficient known at each time step we
computed the magnetospheric response with the parameterized response function
for the successive time step. The equations are non-dimensionalized by
using the magnetic diffusion time scale $\tau = R_\down{cmb}^2/\eta$, the
length-scale $R_\down{cmb}$ and an appropriately chosen magnetic scale $B_0$.
Furthermore, we abbreviate $\tilde{A}(\theta)= A(\theta) \sin\theta$ and
$\tilde{B}(\theta)=B(\theta) \sin\theta$. For the dimensionless
uniform external field we choose
\begin{equation}
   \tilde{A}_\down{ext}(\theta)= \frac{g_{1,\down{ext}}^0}{B_0} \cos \theta \sin
   \theta \quad .
\end{equation}
The $\alpha\Omega$-dynamo equations with an ambient poloidal magnetic
field following the approach by \citet{levy_1979} are written as
\begin{eqnarray}
   \partial_t \tilde{A} &=& \partial^2_\theta \tilde{A} - \cot \theta \;
   \partial_\theta \tilde{A} +
   \cos \theta \tilde{B}
   \label{eq:mod_induction_equation_A}
   \\
   \partial_t \tilde{B} &=& \partial^2_\theta \tilde{B} - \cot \theta \;
   \partial_\theta \tilde{B} + P\sin \theta \left(\partial_\theta \tilde{A} -
   \tilde{A}_\down{ext}\right)
   \label{eq:mod_induction_equation_B}
\end{eqnarray}
with the poloidal external field contributing to the induction effect
acting on the toroidal component. The first two terms of the right-hand side of equations
(\ref{eq:mod_induction_equation_A}) and (\ref{eq:mod_induction_equation_B})
describe the diffusion of the poloidal and toroidal field, respectively, the
third term of (\ref{eq:mod_induction_equation_A}) the action of the
$\alpha$-effect on the toroidal field and the third term of
(\ref{eq:mod_induction_equation_B}) the differential rotation acting on the
internal and the external poloidal field.\\

The model is controlled by the dimensionless dynamo number $P=R_\down{cmb}^4
\Omega' \alpha_0 / \eta^2$. Without an external field the magnetic field grows
when $P$ exceeds a critical value of $P_\down{crit} = 46$ and decays otherwise.
At values $P \gg P_\down{crit}$ the dynamo would begin to show an oscillating
behavior. The complete mode structure is described by
\citet{schmitt_1989}. It is qualitatively also typical for a 2D thick layer
dynamo \citep[see e.g.][]{Parker_1971}. In the present study we are interested
in the monotonically evolving mode for dynamo numbers $P_\down{crit} < P
\lesssim 70$.\\

\begin{figure}
 \centering\includegraphics[width=8cm]{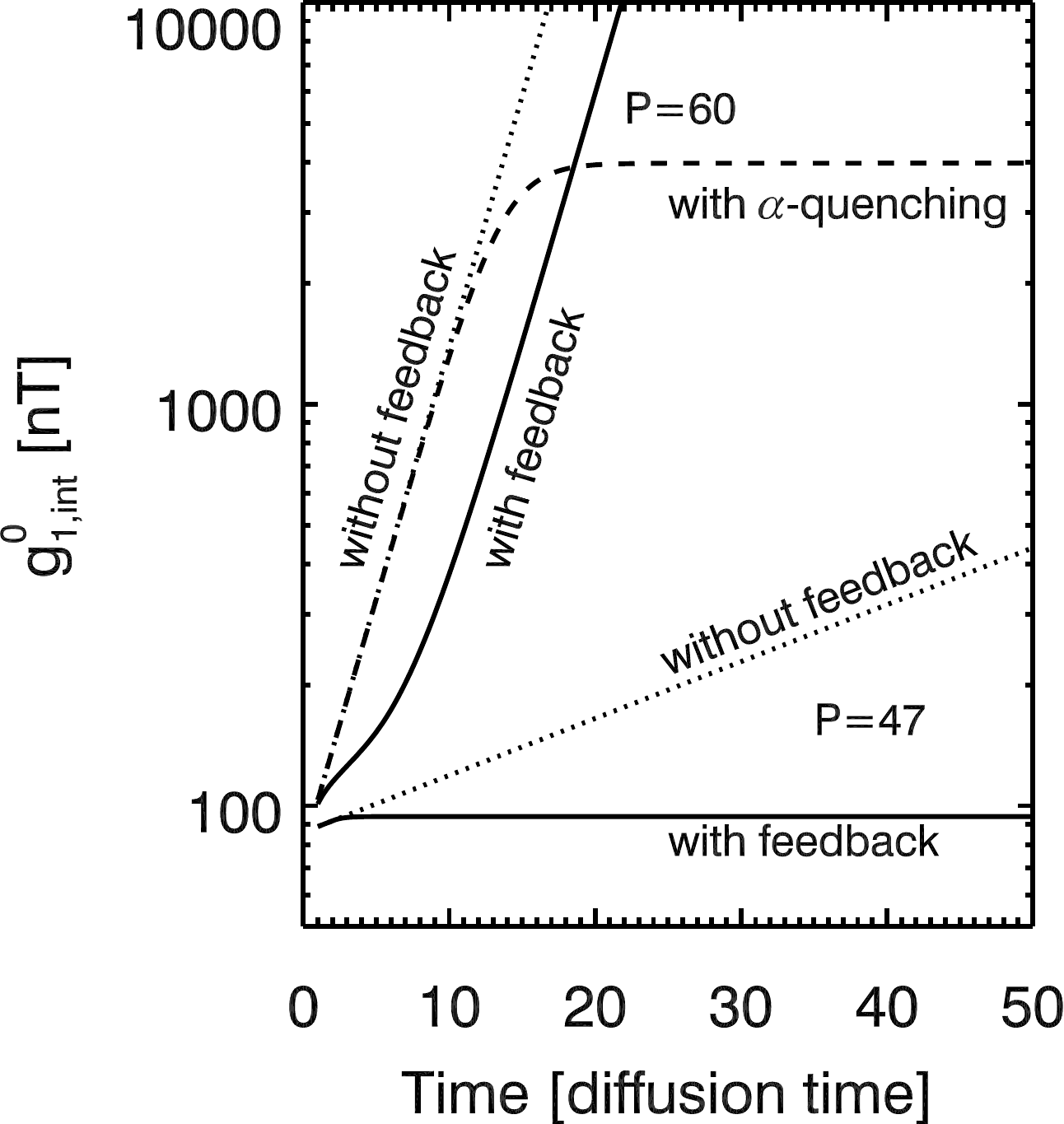}
    \caption{Comparison of the temporal evolution of the internal dipole strength
    depending on different dynamo numbers $P$ and feedback turned on or off.}
    \label{fig:comparison_pure_feedback}
\end{figure}

A toroidal magnetic field of the form $B_\down{seed} = 10^{-5} \sin\theta$ with
a low amplitude compared to $B_0$ serves as seed field in order to avoid a
dependency of the results on the properties of the initial field. Temporal
integration of the modified induction equations
(\ref{eq:mod_induction_equation_A}) and (\ref{eq:mod_induction_equation_B}) is
computed numerically using a finite differencing scheme, where the
diffusion terms are treated implicitly and the induction terms explicitly. The
temporal evolution for different $P$ with and without feedback is presented in
figure \ref{fig:comparison_pure_feedback}. In a fully self-consistent dynamo the
growth of the magnetic field is limited by the Lorentz force acting back on the
flow. Since this feedback mechanism is missing in this kinematic model,
supercritical dynamo numbers $P > P_\down{crit}$ result in an unbounded
exponentially increasing magnetic field strength over time as exemplified for
the two cases in figure \ref{fig:comparison_pure_feedback}. For example,
\citet{schmitt_1989} limit this growth by an $\alpha$-quenching whose effect is
shown as the dashed line in figure \ref{fig:comparison_pure_feedback}.

\begin{figure}
 \centering\includegraphics[width=8cm]{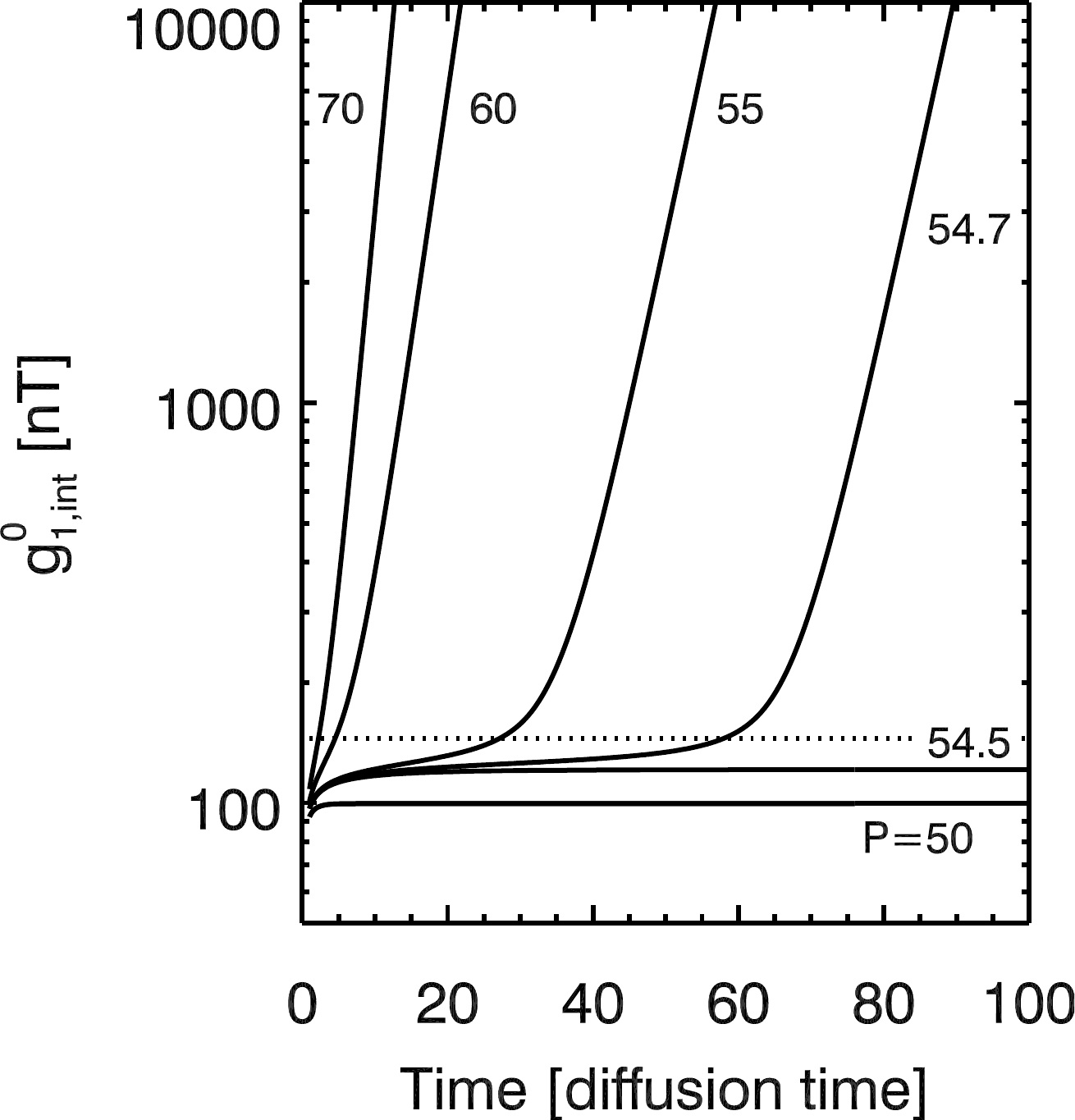}
    \caption{Temporal evolution of the internal dipole strength for various
    dynamo numbers $P$ with the magnetospheric feedback turned on.}
    \label{fig:different_dynamo_numbers}
\end{figure}

The negative feedback from the external field provides an alternative quenching
mechanism. We generally start with a small internal field which is insufficient
to produce a magnetopause above the surface and thus provides no quenching. For
supercritical dynamo numbers the internal field then grows until the external
field has developed sufficiently to provide the necessary quenching as
visualized in figure \ref{fig:comparison_pure_feedback}. This results in a
stationary solution with a magnetic field strength that depends on the dynamo
number.  This can be seen comparing the $P=50$ and $P=54.5$ cases in
figure \ref{fig:different_dynamo_numbers}. The first one is stabilized after about 5
diffusion times, the latter after about 25 diffusion times. The saturation
level is between 100 and 145 nT. However, when the dynamo number is chosen
bigger than $P=54.5$ the quenching is insufficient and the exponential
growth is
only delayed. This happens when the internal dipole strength exceeds
$g_{1,\down{int}}^0=145$ nT where the external field reaches its maximum. This
level is marked in figure \ref{fig:different_dynamo_numbers} with a dotted,
horizontal line. We note that the starting field strength has to be lower than
$145$ nT for the quenching to work.  The saturated field strength is
independent of the initial amplitude. The duration of the delay for $P > 54.5$
and the ultimate exponential growth rate both depend on the dynamo number.  At
$P=54.7$ the exponential growth phase sets in after about 60 diffusion times
whereas the delay has virtually vanished at a dynamo number of $P \geq 70$.

\begin{figure}
 \centering\includegraphics[height=0.25\textheight]{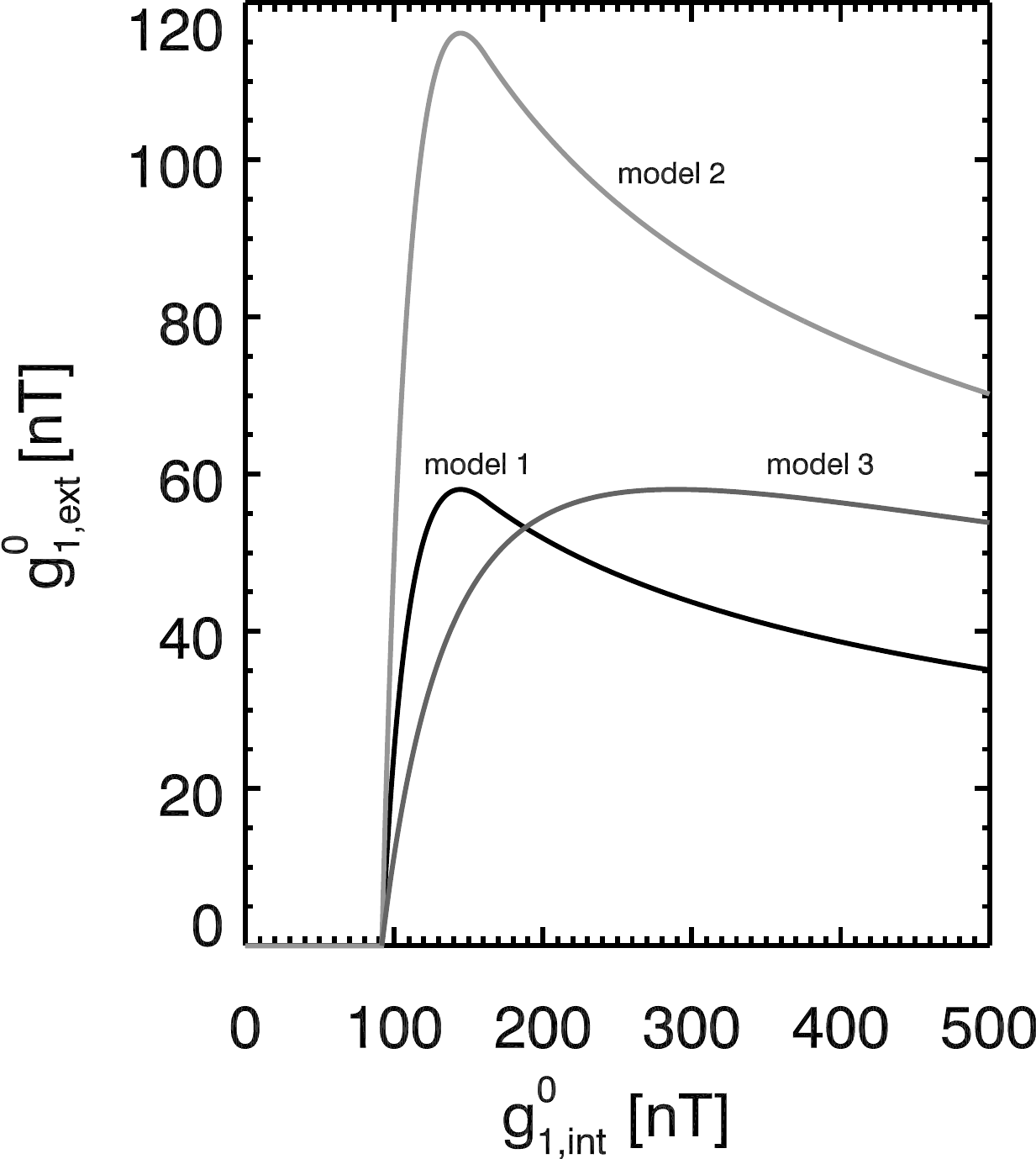}\hspace{0.2cm}
 \includegraphics[height=0.25\textheight]{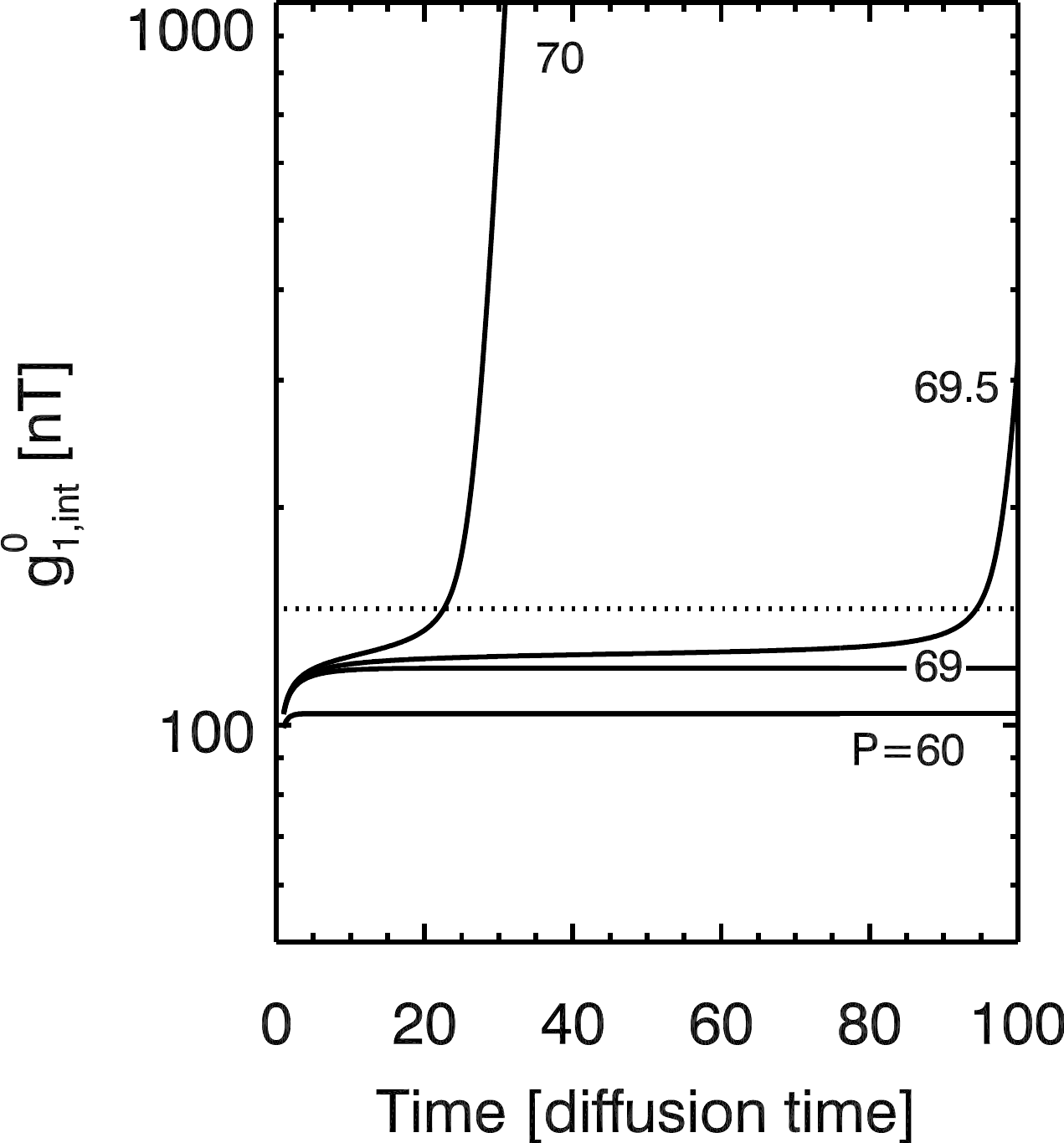}\hspace{0.2cm}
 \includegraphics[height=0.25\textheight]{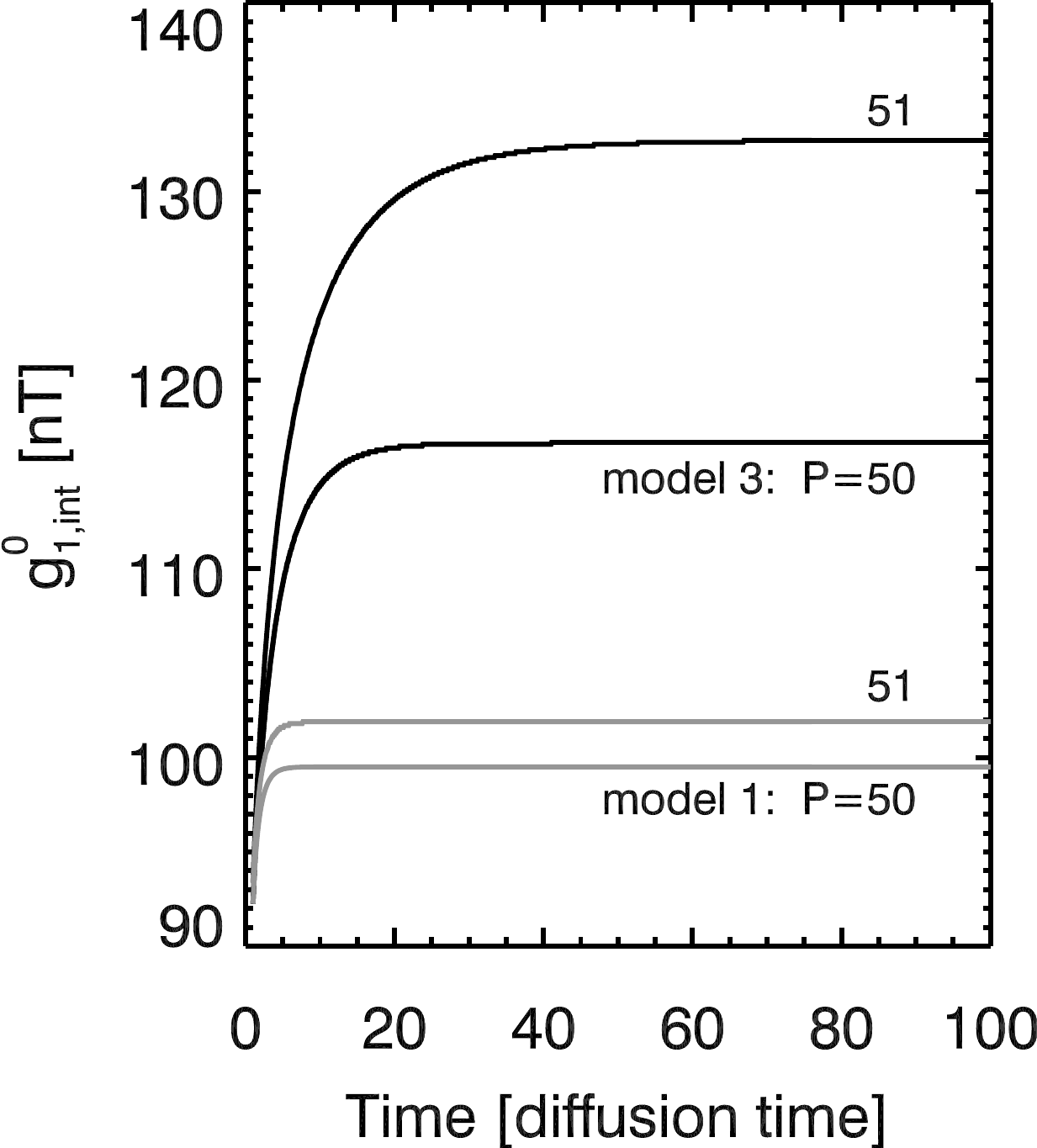}
\caption{Parameter study with artificially altered
response functions. \textit{Left panel}: Model 1 corresponds to the already used form for Mercury.
In model 2 the function of model 1 is amplified by a factor of 2. In model 3
the function has the same amplitude as in model 1 but the maximum is shifted to
a higher internal field value. \textit{Middle panel:} Temporal evolution of the
internal dipole field strength for various dynamo numbers $P$ for model 2.
Compare with figure 4 for model 1. \textit{Right panel:} Stationary saturation
levels of the internal dipole strength for two dynamo numbers $P$ for model
3 in comparison with model 1.}
 \label{fig:altered_response_functions}
\end{figure}

The range of dynamo numbers for which the magnetospheric field limits
the growth of the dynamo, here the relatively small parameter range of
$46\lesssim P\lesssim 54.5$, depends strongly on the maximum value of the
response function. In order to analyze the dependence of the response function,
we investigate two different cases with modified response functions. If we
multiply the response function by a factor of 2 (model 2 in figure
\ref{fig:altered_response_functions}), dynamo numbers up to 69 result in
stationary solutions as is shown in figure \ref{fig:altered_response_functions}.\\

The field strength at which stationary solutions saturate, in the
original model (hereafter referred to as model 1), values between 100 and 145
nT, is determined by the rising part of the response function. If the maximum
is at higher values of $g_\down{1,int}^0$ (model 3 in figure
\ref{fig:altered_response_functions}), a higher saturation field strength
results compared to model 1 for the same dynamo numbers.

\section{Conclusion and outlook}

Using a kinematic $\alpha\Omega$-dynamo in a feedback configuration, we have
demonstrated that the feedback of the external field on the internal dynamo
mechanism can indeed result in relatively small field strengths below $150$ nT
as suggested by \citet{glassmeier_2007}. However, in our simplified kinematic
dynamo model the responsible quenching would only be sufficient in a narrow
regime where the dynamo number does not exceed 18\% of its critical value. If 
Mercury is captured in the quenched regime our model implies that 
the Hermean dynamo is unique. 
It should be noted here that alternative explanations for the
weak Hermean dynamo field \citep[e.g.][]{Stanley_2005,Christensen_2006} also
require the assumption of special conditions for Mercury.
The saturation field strength strongly depends on the assumed response function
describing the dependence of the external field on the internal field strength.
Unfortunately, very little is known about the underlying interaction,
especially for a magnetopause close to the surface which would be appropriate
for Mercury which is neccessary for our suggested feedback mechanism to work.\\

This paper is part of a series of studies examining the model of a feedback
dynamo scenario. \citet{glassmeier_2007} made use of extensively simplified
models and examined stationary dynamo solutions without addressing the question
how these stationary solutions could be realized.  This problem has been
addressed in this study. We further consider an analytical solution
to an approximation of the kinematic dynamo problem which allows us to examine
the influence of the shape of the response function on the dynamo solution. The
results could be useful for the application of the idea of a feedback dynamo to
other astrophysical bodies such as gas giants close to their host star. 
Furthermore, we address the response function (also for higher magnetic
multipoles) for Mercury by using a hybrid code simulating the interaction of
Mercury's magnetosphere with the solar wind. Another investigation concerns how
a three-dimensional, self-consistent, numerical dynamo model in approximate
magnetostrophic balance \citep{wicht_2002} reacts to an imposed uniform and
constant-in-time external field.  From the results of these simulations we
will know what kind of characteristic reactions of the dynamo we can expect
when examining the full time dependent, 3D model with the exact and full
magnetospheric response function.

\section*{Acknowledgement}

\noindent We are grateful to Ulrich Christensen and Natalia Gomez-Perez for
illuminating discussions.  This work was financially supported by the German
Ministerium f\"ur Wirtschaft und Technologie and the German Zentrum f\"ur Luft-
und Raumfahrt under contract 50 QW 0602.



\begin{thebibliography}{21}
\providecommand{\natexlab}[1]{#1}

\bibitem[\protect\citeauthoryear{{Anderson}
  {\itshape{et~al.}}}{2009}]{messenger_2009}
{Anderson}, B.J., {Acu{\~n}a}, M.H., {Korth}, H., {Slavin}, J.A., {Uno}, H.,
  {Johnson}, C.L., {Purucker}, M.E., {Solomon}, S.C., {Raines}, J.M.,
  {Zurbuchen}, T.H., {Gloeckler}, G. and {McNutt}, R.L., {The magnetic field of
  Mercury}. {\itshape Space Sci. Rev.} 2009.

\bibitem[\protect\citeauthoryear{{Baumjohann} and
  {Treumann}}{1996}]{basic_space_plasma_physics_1996}
{Baumjohann}, W. and {Treumann}, R.A., {\itshape {Basic space plasma physics}},
   1996 (London: Imperial College Press).

\bibitem[\protect\citeauthoryear{{Berchem} and {Russell}}{1982}]{russell_1982}
{Berchem}, J. and {Russell}, C.T., {The thickness of the magnetopause current
  layer - ISEE 1 and 2 observations}. {\itshape J. Geophys. Res.} 1982,
  \textbf{87}, 2108--2114.

\bibitem[\protect\citeauthoryear{{Christensen}}{2006}]{Christensen_2006}
{Christensen}, U.R., {A deep dynamo generating Mercury's magnetic field}.
  {\itshape Nature} 2006, \textbf{444}, 1056--1058.

\bibitem[\protect\citeauthoryear{{Glassmeier}}{1997}]{glassmeier_1997}
{Glassmeier}, K.H., {The Hermean magnetosphere and its ionosphere-magnetosphere
  coupling}. {\itshape Planet. Space Sci.} 1997, \textbf{45}, 119--125.

\bibitem[\protect\citeauthoryear{{Glassmeier}
  {\itshape{et~al.}}}{2010}]{bepi_pss_2009}
{Glassmeier}, K.H., {Auster}, H.U., {Heyner}, D., {Okrafka}, K., {Carr}, C.,
  {Berghofer}, G., {Anderson}, B., {Balogh}, A., {Baumjohann}, W., {Cargill},
  P., {Christensen}, U., {Delva}, M., {Dougherty}, M., {Fornacon}, K.H.,
  {Horbury}, T., {Lucek}, E., {Magnes}, W., {Mandea}, M., {Matsuoka}, A.,
  {Matsushima}, M., {Motschmann}, U., {Nakamura}, R., {Narita}, Y.,
  {O’Brien}, H., {Richter}, I., {Schwingenschuh}, K., {Shibuya}, H.,
  {Slavin}, J., {Sotin}, C., {Stoll}, B., {Tsunakawa}, H., {Vennerstrom}, S.,
  {Vogt}, J. and {Zhang}, T., {The fluxgate magnetometer of the BepiColombo
  Mercury planetary orbiter}. {\itshape Planet. Space Sci.} 2010, \textbf{58},
  287--299.

\bibitem[\protect\citeauthoryear{{Glassmeier}
  {\itshape{et~al.}}}{2007}]{glassmeier_2007}
{Glassmeier}, K.H., {Auster}, H.U. and {Motschmann}, U., {A feedback dynamo
  generating Mercury's magnetic field}. {\itshape Geophys. Res. Lett.} 2007,
  \textbf{34}, L22201.

\bibitem[\protect\citeauthoryear{{Heimpel}
  {\itshape{et~al.}}}{2005}]{Heimpel_2005}
{Heimpel}, M.H., {Aurnou}, J.M., {Al-Shamali}, F.M. and {Gomez Perez}, N., {A
  numerical study of dynamo action as a function of spherical shell geometry}.
  {\itshape Earth Planet. Sci. Lett.} 2005, \textbf{236}, 542--557.

\bibitem[\protect\citeauthoryear{{Korth} {\itshape{et~al.}}}{2004}]{korth_2004}
{Korth}, H., {J.~Anderson}, B., {Acu{\~n}a}, M.H., {Slavin}, J.A.,
  {Tsyganenko}, N.A., {Solomon}, S.C. and {McNutt}, R.L., {Determination of the
  properties of Mercury's magnetic field by the MESSENGER mission}. {\itshape
  Planet. Space Sci.} 2004, \textbf{52}, 733--746.

\bibitem[\protect\citeauthoryear{{Levy}}{1979}]{levy_1979}
{Levy}, E.H., {Planetary dynamo amplification of ambient magnetic fields}; in
  {\itshape Lunar and Planetary Science Conference}, edited by N.W. {Hinners},
  Vol. ~10 of {\itshape Lunar and Planetary Science Conference} 1979, pp.
  2335--2342.

\bibitem[\protect\citeauthoryear{{Olson} and
  {Christensen}}{2006}]{olson_christensen_2006}
{Olson}, P. and {Christensen}, U.R., {Dipole moment scaling for
  convection-driven planetary dynamos}. {\itshape Earth Planet. Sci. Lett.}
  2006, \textbf{250}, 561--571.

\bibitem[\protect\citeauthoryear{{Parker}}{1971}]{Parker_1971}
{Parker}, E.N., {The generation of magnetic fields in astrophysical bodies.IV.
  The solar and terrestrial dynamos}. {\itshape Astrophys. J.} 1971,
  \textbf{164}, 491--509.

\bibitem[\protect\citeauthoryear{{Schmitt} and
  {Sch{\"u}ssler}}{1989}]{schmitt_1989}
{Schmitt}, D. and {Sch{\"u}ssler}, M., {Non-linear dynamos. I - One-dimensional
  model of a thin layer dynamo}. {\itshape Astron. Astrophys.} 1989,
  \textbf{223}, 343--351.

\bibitem[\protect\citeauthoryear{{Spohn} {\itshape{et~al.}}}{2001}]{spohn_2001}
{Spohn}, T., {Sohl}, F., {Wieczerkowski}, K. and {Conzelmann}, V., {The
  interior structure of Mercury: what we know, what we expect from
  BepiColombo}. {\itshape Planet. Space Sci.} 2001, \textbf{49}, 1561--1570.

\bibitem[\protect\citeauthoryear{{Stanley}
  {\itshape{et~al.}}}{2005}]{Stanley_2005}
{Stanley}, S., {Bloxham}, J., {Hutchison}, W.E. and {Zuber}, M.T., {Thin shell
  dynamo models consistent with Mercury's weak observed magnetic field}.
  {\itshape Earth Planet. Sci. Lett.} 2005, \textbf{234}, 27--38.

\bibitem[\protect\citeauthoryear{{Suess} and {Goldstein}}{1979}]{suess_1979}
{Suess}, S.T. and {Goldstein}, B.E., {Compression of the Hermaean magnetosphere
  by the solar wind}. {\itshape J. Geophys. Res.} 1979, \textbf{84},
  3306--3312.

\bibitem[\protect\citeauthoryear{{Takahashi} and
  {Matsushima}}{2006}]{Matsushima_2006}
{Takahashi}, F. and {Matsushima}, M., {Dipolar and non-dipolar dynamos in a
  thin shell geometry with implications for the magnetic field of Mercury}.
  {\itshape Geophys. Res. Lett.} 2006, \textbf{33}, L10202.

\bibitem[\protect\citeauthoryear{{Tsyganenko}}{1996}]{Tsy_1996}
{Tsyganenko}, N.A., {Effects of the solar wind conditions in the global
  magnetospheric configurations as deduced from data-based field models}; in
  {\itshape International Conference on Substorms}, edited by E.J. {Rolfe} and
  B.~{Kaldeich}, Vol.  389 of {\itshape ESA Special Publication} 1996, p. 181.

\bibitem[\protect\citeauthoryear{{Tsyganenko} and
  {Sitnov}}{2005}]{Tsyganenko_2005}
{Tsyganenko}, N.A. and {Sitnov}, M.I., {Modeling the dynamics of the inner
  magnetosphere during strong geomagnetic storms}. {\itshape J. Geophys. Res.}
  2005, \textbf{110}, A3208.

\bibitem[\protect\citeauthoryear{{Wicht}}{2002}]{wicht_2002}
{Wicht}, J., {Inner-core conductivity in numerical dynamo simulations}.
  {\itshape Phys. Earth Planet. Inter.} 2002, \textbf{132}, 281--302.

\bibitem[\protect\citeauthoryear{{Wicht} {\itshape{et~al.}}}{2007}]{wicht_2007}
{Wicht}, J., {Mandea}, M., {Takahashi}, F., {Christensen}, U.R., {Matsushima},
  M. and {Langlais}, B., {The origin of Mercury's internal magnetic field}.
  {\itshape Space Sci. Rev.} 2007, \textbf{132}, 261--290.

\end{thebibliography}

\end{document}